\documentclass{sig-alternate-05-2015}
\usepackage{graphicx}
\usepackage{caption}
\usepackage{subcaption}

\begin{document}

\CopyrightYear{2016} 
\setcopyright{acmlicensed}
\conferenceinfo{MSR'16,}{May 14-15 2016, Austin, TX, USA}
\isbn{978-1-4503-4186-8/16/05}\acmPrice{\$15.00}
\doi{http://dx.doi.org/10.1145/2901739.2901767}

\title{From Query to Usable Code: An Analysis of Stack Overflow Code Snippets}
%
%
%
%
%

\numberofauthors{3} 
%
\author{
%
%
\alignauthor
Di Yang\\
       \affaddr{Department of Informatics}\\
       \affaddr{University of California, Irvine}\\
       \email{diy4@uci.edu}
\alignauthor
Aftab Hussain\\
       \affaddr{Department of Informatics}\\
       \affaddr{University of California, Irvine}\\
       \email{aftabh@uci.edu}
\alignauthor
Cristina Videira Lopes\\
       \affaddr{Department of Informatics}\\
       \affaddr{University of California, Irvine}\\
       \email{lopes@uci.edu}
}

\date{21 January 2016}

\maketitle
\begin{abstract}

Enriched by natural language texts, Stack Overflow code snippets are
an invaluable code-centric knowledge base of small units of
source code. Besides being useful for software developers, these
annotated snippets can potentially serve as the basis for automated
tools that provide working code solutions to specific natural language
queries. 

With the goal of developing automated tools with the Stack Overflow
snippets and surrounding text, this paper investigates the following
questions: (1) How usable are the Stack Overflow code snippets? and
(2) When using text search engines for matching on the natural
language questions and answers around the snippets, what percentage of
the top results contain usable code snippets?

A total of 3M code snippets are analyzed across four languages: C\#,
Java, JavaScript, and Python. Python and JavaScript proved to be the
languages for which the most code snippets are usable. Conversely,
Java and C\# proved to be the languages with the lowest usability
rate. Further qualitative analysis on usable Python snippets shows
the characteristics of the answers that solve the original question. Finally,
we use Google search to investigate the alignment of
usability and the natural language annotations around code snippets, and
explore how to make snippets in Stack Overflow an
adequate base for future automatic program generation.

\end{abstract}

%
%

\ccsdesc[100]{Software and its engineering~Software libraries and repositories}

%
%

%
%
\printccsdesc


\keywords{code mining, automatic program generation}

\section{Introduction}
\label{sec:Intro}

Research shows that programmers use web searches extensively to look
for suitable pieces of code for reuse, which they adapt to their
needs~\cite{sim-lopes:oss08,Gallardo-Valencia:2009,Philip:cscw2012}. Among
the many good sites for this purpose, Stack Overflow (SO, from here
onwards) is one of the most popular destinations in Google search
results. Over the years, SO has accumulated an impressive amount of
programming knowledge consisting of snippets of code together with
relevant natural language explanations. Besides being useful for
developers, SO can potentially be used as a knowledge base for tools
that automatically combine snippets of code in order to obtain more
complex behavior. Moreover those more complex snippets could be
retrieved by matches on the natural language (i.e. non-coding
information) that enriches the small snippets in SO.

\begin{figure}
\centering
\includegraphics[width=0.5\textwidth]{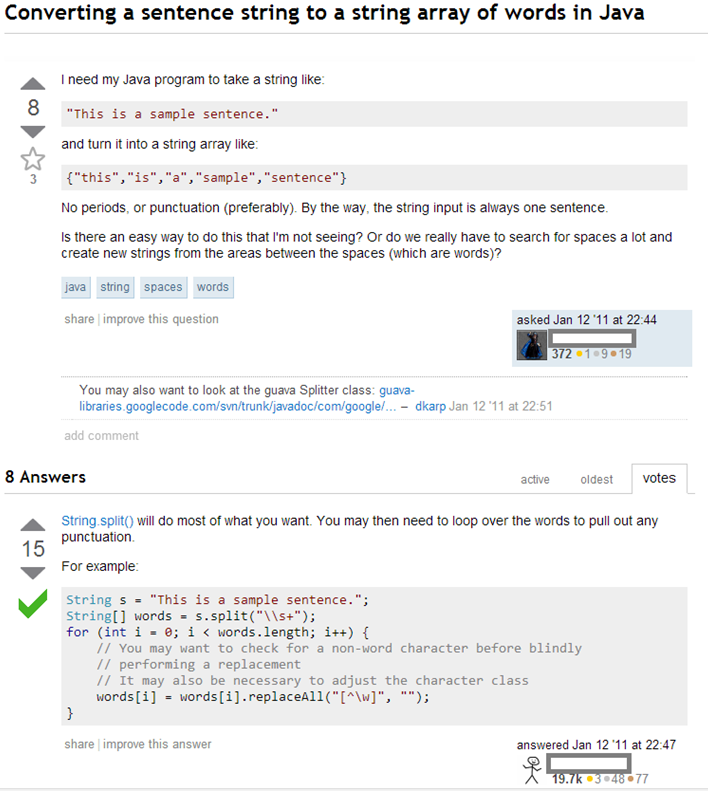}
\caption{Example of Stack Overflow question and answers. The
  search query was ``java parse words in string''. }
\label{fig:so-question}
\end{figure}

As an illustrative example, consider searching for ``java parse words
in string'' using Google Web search. This yields several results in
SO, one of which is shown in Figure~\ref{fig:so-question}.  The
snippet of code provided with the answer that was accepted is almost
executable as-is by copy-paste. The job of programmers becomes, to a
large extent, to glue together these snippets of code that do some
generic functionality by themselves. The fact that a Web search engine
returned this SO page as one of the top hits for our query closes in
on one of the hardest parts of program synthesis, namely the
expression of complex program specifications. Hence, it is conceivable
that tools might be developed that would do that gluing job
automatically given a high-level specification in natural language.

In pursuing this goal, the first challenge one faces is whether, and
to what extent, the existing snippets of code that are suggested by
Web search results are {\em usable} as-is. If there are not enough
usable snippets of code, the process of repairing them automatically
for further composition may be out of reach. This paper presents
research in this direction, by showing the results of our
investigation of the following questions: 
\begin{itemize}
\item [(1)] How usable are the SO code snippets? 
\item [(2)] When using Web search engines for matching on
the natural language questions and answers around the snippets, what
percentage of the top results contain usable code snippets?
\end{itemize}

In order to compare the {\em usability} of different pieces of code,
we need to define what {\em usability} is in the first place. We
classified snippets of code based on the effort that would
(potentially) be required by a program generation tool to use the
snippet as-is. Usability is therefore defined based on the standard
steps of parsing, compiling and running source code. For each of these
steps, if the source code passes, the more likely it is that the tool
can use it with minimum effort.

Given this definition of usability, there are situations where a
snippet that does not parse is more useful than the one that runs, but
passing these steps assures us of important characteristics of the
snippet, such as the code being syntactically and structurally
correct, or all the dependencies being readily available, which are of
surmount importance for automation.

We first study the percentages of parsable, compilable and runnable
(where these steps apply) snippets for each of the four most popular
programming languages (C\#, Java, JavaScript, and
Python).\footnote{Based on the RedMonk programming language popularity
  rankings as of January 2015, four of the most popular
  programming languages are Java, C\#, JavaScript and Python. We
  choose these four, also as representatives of statically-typed (the
  first two) and dynamically-typed languages (the last two).}  From
the results, we saw a significant difference in repair effort
(usability) between the statically-typed the dynamically-typed
languages, the latter being far less effort. Next, we focused on the
best performing language (Python) and conducted a 3-step qualitative
analysis to see if the runnable snippets can actually answer questions
correctly and completely. Finally, in order to close the circle, we
use Google Search in order to find out the extent to which the SO
snippets suggested by the top Google results are usable. Being able to
find a large percentage of usable snippets among the top search
results for informal queries, the idea of automating snippet repair and
composition, and finding those synthetic pieces of code via informal
Web queries becomes within the realm of possibility.

The remainder of this paper is organized as follows. In
Section~\ref{sec:usabilityRates}, we present the overall research
methodology and environment of our work. The results of qualitative
analysis are explained in Section~\ref{sec:qualitative}. In
Section~\ref{sec:googleResults} we investigate the usability and
quality of top results from Google Web search. In Section~\ref{sec:back},
we present the related work in the areas of SO analyses, enhancing
coding environments, and automated code
generation. Section~\ref{sec:concl} concludes the paper.

\section{Usability Rates}
\label{sec:usabilityRates}

This section describes our usability study of SO snippets. In
\ref{subsec:goal}, we elaborate upon our goal. In
\ref{subsec:snippExtrac}, we describe the characteristics of the
extracted snippets. In \ref{subsec:AnalyzingSnippetExec}, we present
the operations that were carried out on the snippets of each
language. We also describe the libraries that were used to process the
snippets for each of the languages, and highlight the limitations
found\footnote{Note to reviewers: code and data for this study are
  available upon request, and will be made publicly available upon
  publication of this paper.}. In \ref{subsec:findings} and
\ref{subsec:errMsg}, we present the usability rates and error messages
for each language.
ƒ
\subsection{Goal}
\label{subsec:goal}
Our goal is to compare the usability rates for the snippets of four programming languages (C\#, Java, JavaScript, and Python) as they exist in SO, i.e., in small snippets of code.
We also want compare the languages regarding their static or dynamic nature, and target the most usable language.
In our study, snippets in Java
and C\#, which are statically-typed, are parsed
and compiled in order to assess their usability level. JavaScript does
not have the process of compilation, so we investigate only the
parsing and running levels for it. Python is also a dynamic language
but it can be compiled. However, this step is not as important as it is for Java and C\#, as important errors (such as name
bindings) are not checked at compile time. As such, for Python, 
like for JavaScript, we assess usability only by parsing and running the snippets.

\subsection{Snippets Extraction}
\label{subsec:snippExtrac}

All snippets were extracted from the dump available at the Stack Exchange data dump
site\footnote{https://archive.org/details/stackexchange, obtained on April 2014.}.

In SO both questions and answers are considered as \textit{posts},
which are stored with unique ids in a table called \texttt{Posts}. In this
table, posts that represent questions and answers are
distinguished by the field \texttt{PostTypeId}. Information about posts can be
obtained by accessing the field \texttt{Body}.

There are two types of answers in SO, \textit{accepted answer} and \textit{best answer}. An accepted answer is the answer
chosen by the original poster of the question to be the most suitable. All question posts
have an \texttt{AcceptedAnswerId} field from which we can identify
accepted answers when these exist. The best answer is the one which has the most number of
votes from other SO users. The vote count is stored in the field \texttt{ViewCount} of the table \texttt{Posts}. Thus, an accepted answer
may not always be the best answer. 

Finally, in SO, questions are tagged with their subject areas, which include relevant information such as
the language or the domain where the question is relevant (networking, text processing, etc.). We get the language information for each accepted answer
from these tags, one example on how the social nature of SO helps categorizing and selecting pieces of code.

In this work, we only include snippets found in all \textbf{accepted answers}. We choose accepted these as we value the 
agreement from the original poster, accepting the fact that it is very likely this answer resolved the original problem. For all posts
for a language we were interested in, we used the the markdown \texttt{<code>} to  extract the code snippets from the field \texttt{Body}.

\subsection{Snippets Processing}
\label{subsec:AnalyzingSnippetExec}

In Table~\ref{tab:ops} we present the operations we performed to
analyze and rate each of language. All the snippets from all languages were parsed,
but depending on the static or dynamic nature of the language we either compiled it
and analyzed the (possible) errors, or ran the language (below we detail these processes).

\begin{table}[h]
\caption{Operations performed for each language.}
\label{tab:ops}
\centering
\begin{tabular}{|l|c|c|c|c|}
\hline
\textbf{Operation} & \multicolumn{1}{l|}{C\#} & \multicolumn{1}{l|}{Java} & \multicolumn{1}{l|}{JavaScript} & \multicolumn{1}{l|}{Python} \\ \hline
Parse              & x                        & x                         & x                               & x                           \\ \hline
Compile            & x                        & x                         &                                 &                             \\ \hline
Run                &                          &                           & x                               & x                           \\ \hline
\end{tabular}
\end{table}

\begin{figure}
	\centering
	\includegraphics[scale = 0.35]{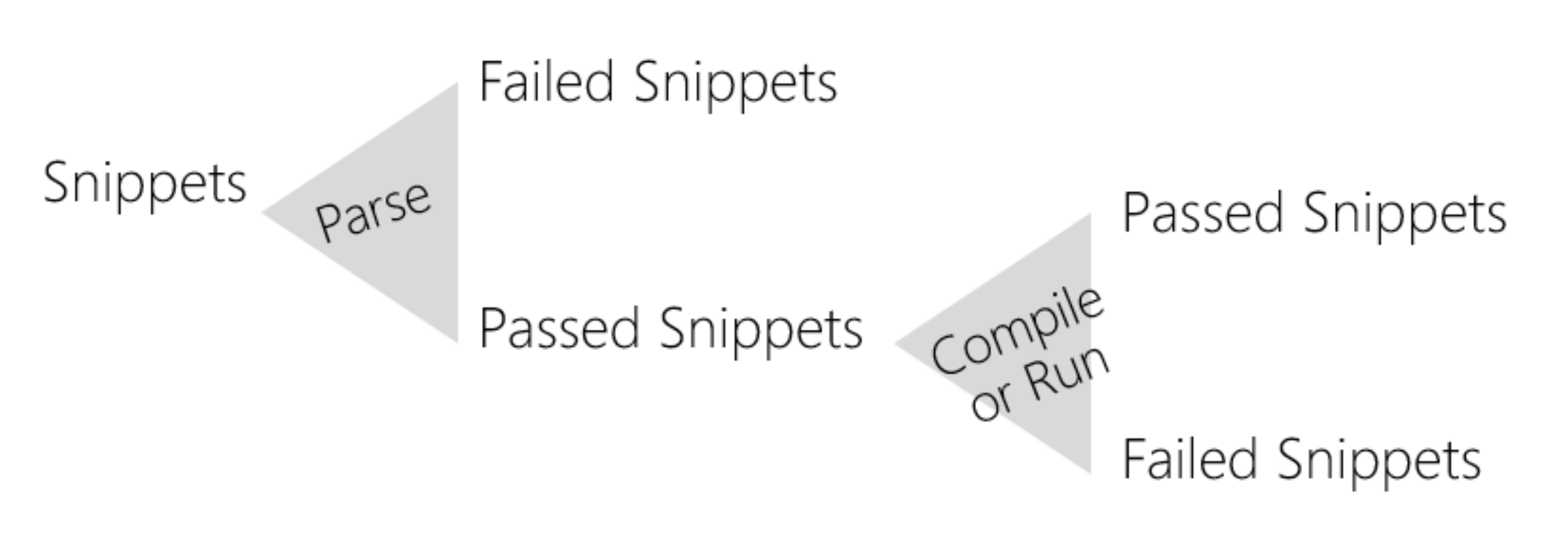} 
	\caption{Sequence of operations}
	\label{fig:opseq}
	\end{figure}

Figure~\ref{fig:opseq} shows the order in which these operations
are performed. We compile (or run) only those snippets
which passed parsing, since snippets which are unparsable have syntactic errors and therefore are also
non-compilable/non-runnable.

We used a set of tools and APIs to process the snippets in the various languages, which we present next:

\subsubsection{C\#}
For parsing we utilize a tool called Roslyn by Microsoft to
obtain the parsable snippets. Roslyn provides for the Visual
Studio languages rich APIs for different phases of compilation. In
particular, Roslyn provides us with the API for getting the abstract
syntax tree, which is the landmark of parsing process. Syntax errors
will be detected in this step.

Compiling C\# programmatically can be easily done by using a
functionality provided by the .NET Framework and found in the
\texttt{Microsoft.CSharp} and \texttt{System.CodeDom.Compiler}
namespaces. We need to call the function that compiles the code, and results
of whether a snippet compiles or not are returned together with errors if applicable.

\subsubsection{Java}
Eclipse's JDT (Java Development Tools) Parser (ASTParser) and the
Javax.Tools were used for the parsing and compiling processes,
respectively. We use JDT1.7 in our experiments since it's the latest version 
for our data dump.

Using the JDT's \texttt{ASTParser} we generated abstract
syntax trees of the snippets, and any parse errors found during the process were extracted via
the \texttt{IProblems} interface.

Javax.Tools compilation functionality first creates a
dynamic source code file object of the Java snippet, from which it
generates a list of compilation units, which are passed as parameters
to the \texttt{CompilationTask} object for compilation. Issues
during compilation are stored in a \texttt{Diagnostics} object. Issues
could be of the following kinds: \texttt{ERROR, MANDATORY\_WARNING,
  NOTE, OTHER, WARNING}. We only look for issues which are of kind
\texttt{ERROR}, as they are the ones more likely prevent the
normal completion of compilation.

\subsubsection{JavaScript}
A reflection of the SpiderMonkey parser is included in the
SpiderMonkey JavaScript Shell and is made available as a JavaScript API. It
parses a string as a JavaScript program and returns a \texttt{Program}
object representing the parsed abstract syntax tree. Syntax errors
are thrown if parsing fails. JavaScript Shell has also a built-in
function \texttt{eval()} to execute JavaScript code, which we used if parsing succeed.

A limitation of the SpiderMonkey parser is that it terminates the
processing of a snippet right when it encounters the first
error. Therefore, it does not identify {\em all} the errors in a
snippet, only the first one, but this suffices to detect problems in the code.

\subsubsection{Python}
Python's built-in AST module and \texttt{compile()} method can help us parse code strings.
Python is a special language among dynamic languages: it has the process of building abstract syntax tree into Python code objects,
so it has the \texttt{compile} function. But when we specify one of the function parameters to be AST only, it only parse the code
by building the AST. \texttt{exec} statement provides functionality to run code strings.

One problem we encountered in processing Python snippets was that Python2 and Python3 have
some incompatible language features. To deal with snippets written in
different versions of Python, and to avoid being biased when rating these pieces of code, we first examined all Python snippets
under Python2 engine, and examined the unparsable ones again under the
Python3 engine and combine the results. The Python libraries share the same limitation as JavaScript's SpiderMonkey; they do not catch {\em all} the errors in a
snippet, only the first one.

\subsection{Findings}
\label{subsec:findings}
 We present the results that were obtained after the
 initial parsing and compiling (or running) of the snippets.

\begin{figure}
\centering
\includegraphics[width=0.5\textwidth]{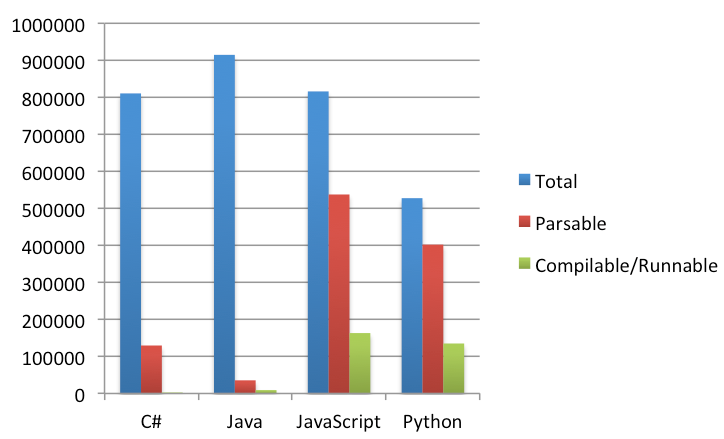} 
\caption{Parsable and compilable/runnable rates histogram}
\label{fig:ini_res_pic}
\end{figure}

Table~\ref{tab:ini_res} and Figure~\ref{fig:ini_res_pic} shows the summary of usability results of all
the snippets. A total of 3M code snippets were analyzed. Python and JavaScript proved to be the
languages for which the most code snippets are usable: 537,767 (65.88\%)
JavaScript snippets are parsable and 163,247 (20.00\%) of them are
runnable; for Python, 402,249 (76.22\%) are parsable and 135,147 (25.61\%)
are runnable. Conversely, Java and C\# proved to be the languages with
the lowest usability rate: 129,727 (16.00\%) C\# snippets are parsable
but only 986 (0.12\%) of them are compilable; for Java, only 35,619
(3.89\%) are parsable and 9,177 (1.00\%) compile.

\begin{table*}[!htbp]
\centering
\caption{Summary of results}
\label{tab:ini_res}
\begin{tabular}{lllll}
                                          & \textbf{C\#}      & \textbf{Java}      & \textbf{JavaScript} & \textbf{Python}      \\ \hline
\textbf{Total Snippets Processed}         & 810,829           & 914,974         & 816,227             & 527,774              \\ \hline
\textbf{Parsable Snippets (Percentage)}   & 129,727 (16.00\%) & 35,619 (3.89\%) & 537,767 (65.88\%)   & 402,249 (76.22\%)\\ \hline
\textbf{Compilable Snippets (Percentage)} & 986 (0.12\%)      & 9,177(1.00\%)   & --                  & --\\ \hline
\textbf{Runnable Snippets (Percentage)}   & --                & --              & 163,247 (20.00\%)   & 135,147 (25.61\%)                
\end{tabular}
\end{table*}

As a result of finding such low parsable and compilable rates for Java and C\#, 
we removed Java and C\# snippets that only contained single words
(i.e. tokens without non-alphabetical characters). The rationale behind this step
was to that a single word in C\# or Java is too insignificant a
candidate for composability; by ignoring those snippets we might
improve the usability rates for these two languages. We then parsed
and compiled the remaining snippets, the results of which are shown in
Table~\ref{tab:ini_res_postremv}. We see that the rates of usability
improve for both languages, and for both parsing and compilation. 
For Java, the parsable rate increases from 3.89\% to 6.22\%, and the compilable rate increases from
1.00\% to 1.60\%. For C\#, the parsable rate increases from 16.00\% to 25.18\%, and the compilable rate
increases from 0.12\% to 0.19\%.

\begin{table}[!htbp]
\centering
\caption{Summary of results for C\# and Java after single-word
  snippets removal}
\label{tab:ini_res_postremv}
\begin{tabular}{lll}
                                                & \textbf{C\#}      & \textbf{Java}      \\ \hline
\begin{tabular}[c]{@{}l@{}}\textbf{Total snippets}\\      \textbf{after removal}\end{tabular}           & 514,992           & 572,742\\ \hline
\textbf{Parsable}         & 129,691 (25.18\%) & 35,619 (6.22\%) \\ \hline
\textbf{Compilable}       & 986 (0.19\%)      & 9,177 (1.60\%) 
\end{tabular}
\end{table}

\subsection{Snippet Examples}
\begin{figure}
\centering
\includegraphics[width = 0.5\textwidth]{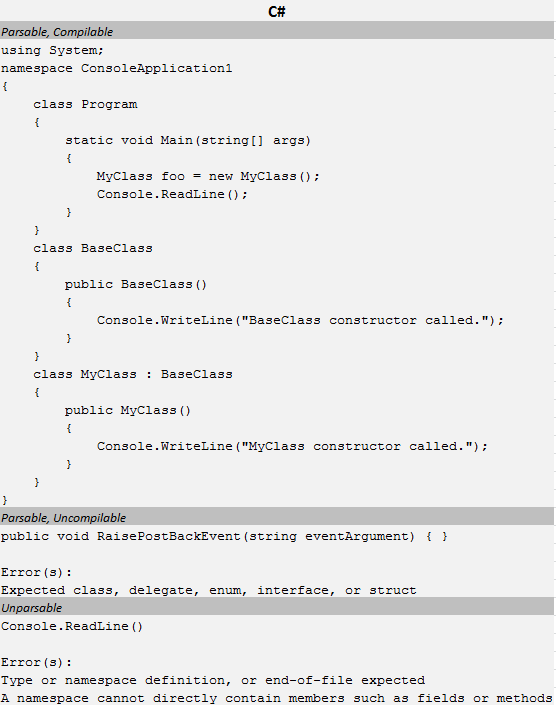}
\caption{Examples of C\# Snippets.}
\label{fig:Cshrpexamples}
\end{figure}

\begin{figure}
\centering
\includegraphics[width = 0.5\textwidth]{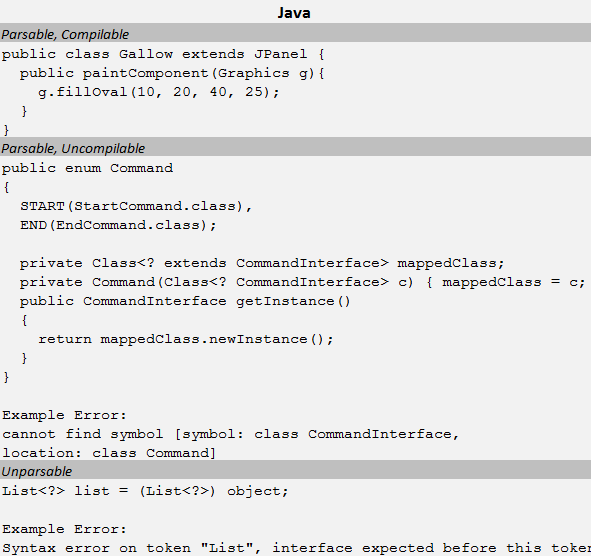}
\caption{Examples of Java Snippets.}
\label{fig:Javaexamples}
\end{figure}

\begin{figure}
\centering
\includegraphics[width = 0.35\textwidth]{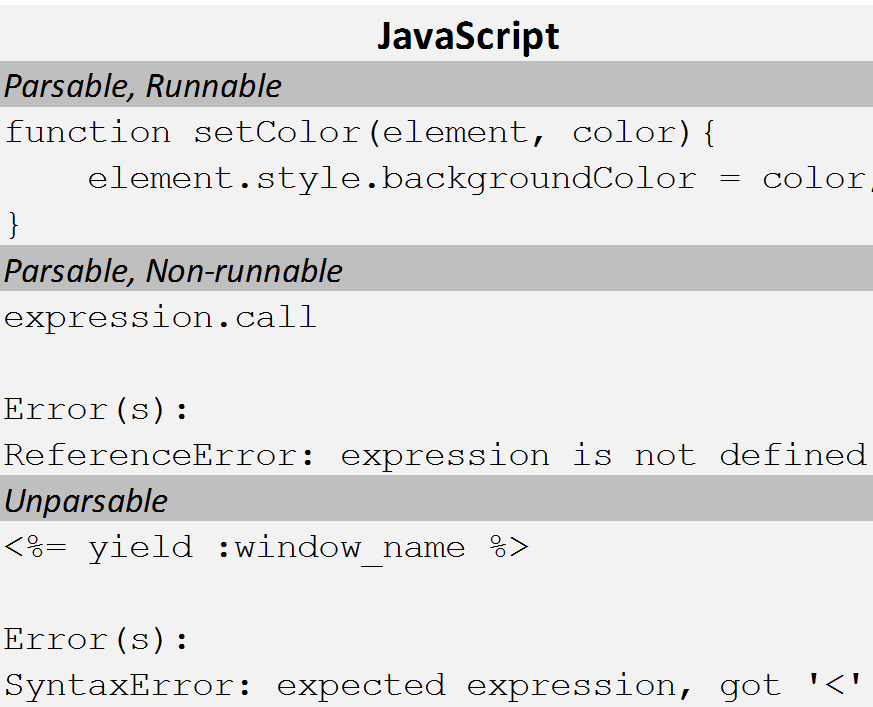}
\caption{Examples of JavaScript Snippets.}
\label{fig:JSexamples}
\end{figure}

\begin{figure}
\centering
\includegraphics[width = 0.35\textwidth]{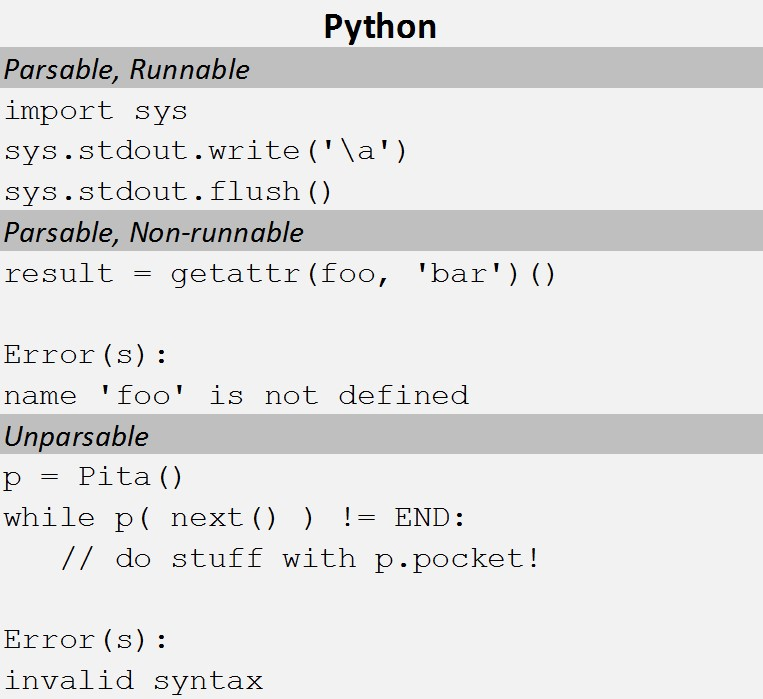}
\caption{Examples of Python Snippets.}
\label{fig:Pythexamples}
\end{figure}

Figures~\ref{fig:Cshrpexamples}, \ref{fig:Javaexamples},
\ref{fig:JSexamples}, and~\ref{fig:Pythexamples} show examples of SO
snippets that passed and failed for each language. For those that
failed, we also show the generated error messages. The examples are representative of the most common error messages. 

The unparsable
Python snippet in Figure~\ref{fig:Pythexamples} also illustrates a
common occurrence in SO posts, where the example code is given more or
less as pseudo-code that mixes the syntax of several languages.


\subsection{Error Messages}
\label{subsec:errMsg}
During the usability analysis process, we log the common errors of the four languages.
In total, we collected 3,347,674 parse errors and 359,783 compile errors for C\#, 
1,417,910 parse errors and 199,489 compile errors for Java,
278,460 parse errors and 374,520 runtime errors for JavaScript,
and 125,525 parse errors and 267,102 runtime errors for Python.

\begin{figure*}
        \centering
        \begin{subfigure}[b]{0.5\textwidth}
	\includegraphics[width=\textwidth]{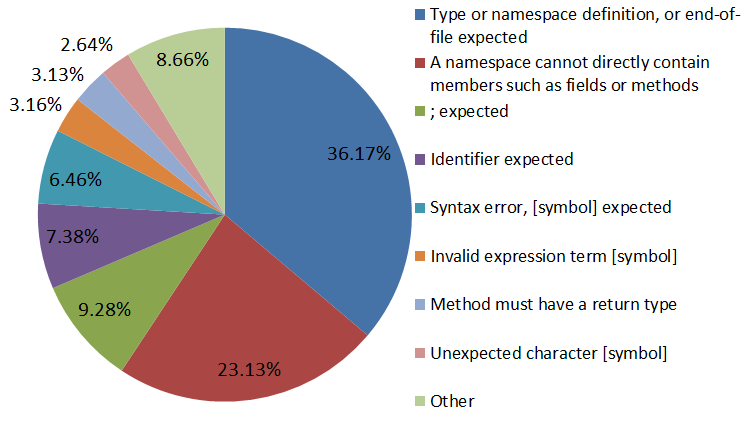}
	\caption{Most common parse errors for C\# }
	\label{fig:CshrpParseErr}
        \end{subfigure}%
        \begin{subfigure}[b]{0.5\textwidth}
	\includegraphics[width=\textwidth]{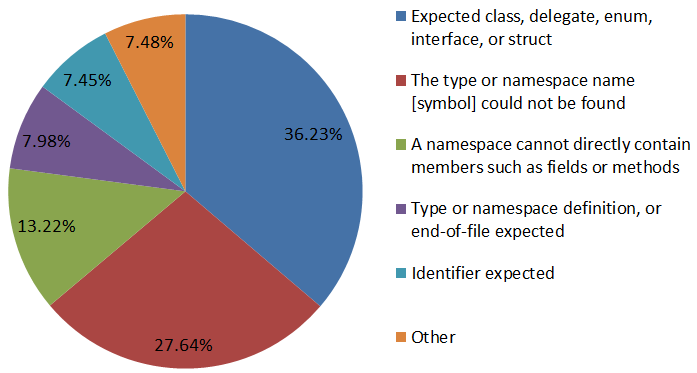}
	\caption{Most common compile errors for C\# }	
	\label{fig:CshrpCompileErr}
        \end{subfigure}
        \caption{Most common error messages for C\#}
        \label{fig:CshrpError}
\end{figure*}

\begin{figure*}
        \centering
        \begin{subfigure}[b]{0.5\textwidth}
        	\centering
	\includegraphics[width=\textwidth]{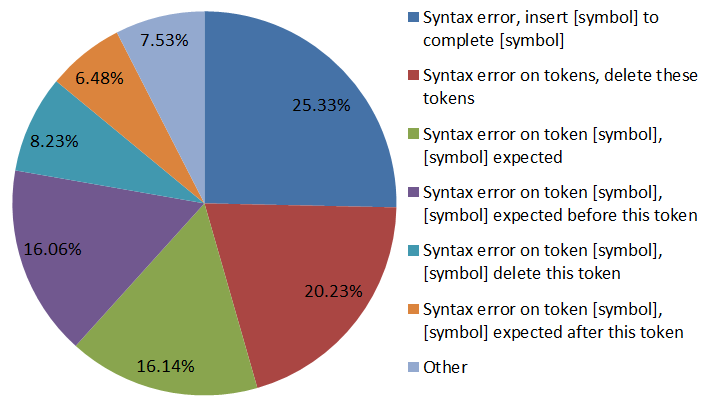}
	\caption{Most common parse errors for Java }
	\label{fig:JavaParseErr}
        \end{subfigure}%
        \begin{subfigure}[b]{0.5\textwidth}
        	\centering
	\includegraphics[width=\textwidth]{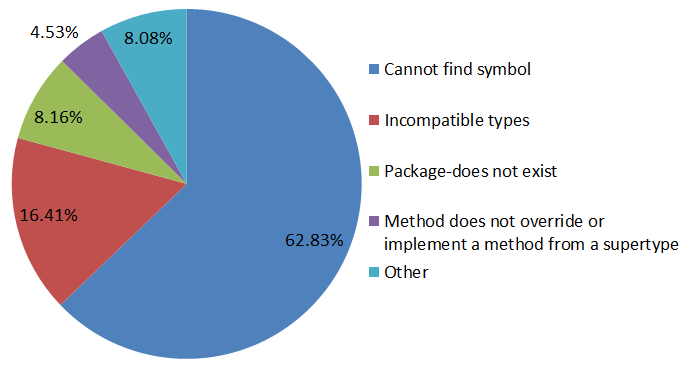}
	\caption{Most common compile errors for Java}
	\label{fig:JavaCompileErr}
        \end{subfigure}
        \caption{Most common error messages for Java}
        \label{fig: JavaError}
\end{figure*}

\begin{figure*}
        \centering
        \begin{subfigure}[b]{0.5\textwidth}
        	\centering
	\includegraphics[width=\textwidth]{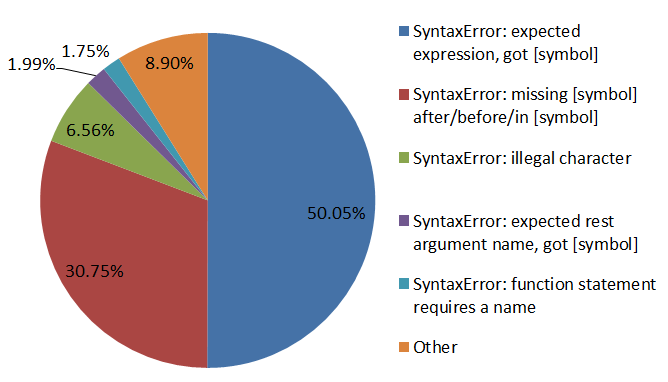}
	\caption{Most common parse errors for JavaScript }
	\label{fig:JSParseErr}
        \end{subfigure}%
        \begin{subfigure}[b]{0.5\textwidth}
        	\centering
	\includegraphics[width=\textwidth]{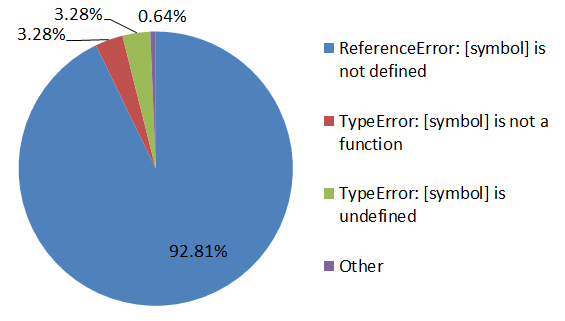}
	\caption{Most common runtime errors for JavaScript }
	\label{fig:JSRunErr}
        \end{subfigure}
        \caption{Most common error messages for JavaScript}
        \label{fig: JSError}
\end{figure*}

\begin{figure*}
        \centering
        \begin{subfigure}[b]{0.5\textwidth}
        	\centering
	\includegraphics[width=\textwidth]{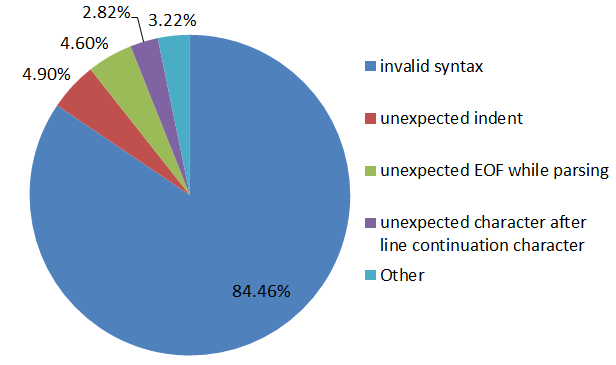}
	\caption{Most common parse errors for Python }
	\label{fig:PythonParseErr}
        \end{subfigure}%
        \begin{subfigure}[b]{0.5\textwidth}
        	\centering
	\includegraphics[width=\textwidth]{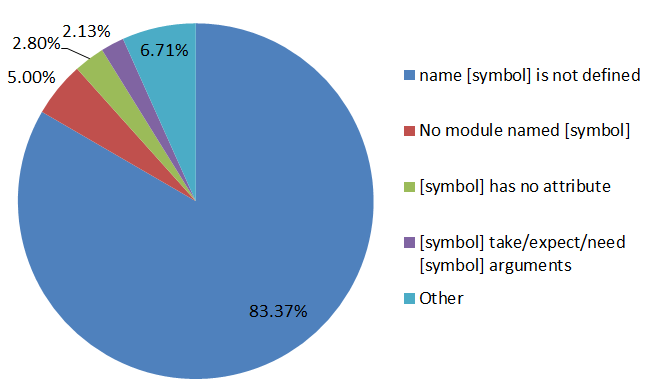}
	\caption{Most common runtime errors for Python }
	\label{fig:PythonRunErr}
        \end{subfigure}
        \caption{Most common error messages for Python}
        \label{fig: PythonError}
\end{figure*}

The common error messages for C\# are shown in
Figure~\ref{fig:CshrpParseErr} and~\ref{fig:CshrpCompileErr}, for Java in
Figure~\ref{fig:JavaParseErr} and~\ref{fig:JavaCompileErr},
for JavaScript in Figure~\ref{fig:JSParseErr} and~\ref{fig:JSRunErr}, and for Python in
Figure~\ref{fig:PythonParseErr} and~\ref{fig:PythonRunErr}. 
The error messages are listed in descending order of percentage. 
The token `[symbol]' is just a replacement for various specific strings appeared in error messages.

The error messages
shown for Java and C\# were obtained on the collection of snippets
without single-words. 
It is important to note that the libraries used
for Python and JavaScript can generate at most one error message for a
snippet, so they do not show all problems that each snippet may have.

Main syntax problems are shown in parsing errors, for all four languages. For example for JavaScript, 50\% of the parsing errors are not getting an expression. For Java, 25\% of the parsing errors are tokens to be inserted.

Errors more related to code context, such as missing symbols, are revealed in compiling or running process.
For C\#, ``type or namespace'' is a main issue for usability. For Java, ``cannot find symbol'' dominates compiling error messages.
When running a JavaScript snippet, we are most likely to stop at a reference error, while for Python, the most common runtime
error is a specific name not defined. 

The purpose for logging the error messages is to provide a knowledge base for repairing codes and increasing usability rates in the future.
For example one of the main parsing errors for C\# is missing semicolons, then a heuristic repair to C\# codes to improve parsable rate 
can be locating missing semicolons and append them. In next section, we give example of two heuristic repairs for Java and C\# snippets.

\subsection{Heuristic Repairs for Java and C\# Snippets}
\label{subsec:fixes}

From the preliminary results above, we can see that the parsing rates
for Python and JavaScript are significantly better than Java and
C\#. The parsing errors reveal the main syntax problems, while the compiling errors given above are more related to code context, such as missing symbols. In
this case, compiling errors are hard to fix, because we need to look
into the specific snippet to complement the missing symbols.

Based on the common error messages for Java and C\#, we implemented two
heuristic repairs on Java and one repair on C\# snippets in order to improve their parsing and compilation
rates.

\subsubsection{Repair 1 - Class}

Many Java snippets consist just of Java code without it being properly
encapsulated in a class or a method. The \texttt{class} construct is
essential for Java snippets. The
\texttt{class} repair fixes Java code snippets that were found to be
missing a class construct based on a heuristic check. This heuristic
check works as follows: if the code snippet is found to contain any of
the tokens \texttt{import}, \texttt{package}, or \texttt{class}, we
assume that the \texttt{class} construct already exists in the
snippet. The rationale behind this heuristic is that, based on our
observations of the snippets, tokens \texttt{import} and
\texttt{package} form scaffolding information of code in SO and are
not the focus of SO answers. Hence any code that uses one of them is
likely to use the \texttt{class} construct also. We also assume if
a token \texttt{class} is present in a code, it exists with enclosing
braces and as a keyword and not a part of a string or comment.

Example:
\begin{verbatim}
\\Repair 1 Candidate
public void main(String args []){
  System.out.println("Hello World");
}
\\After Repair 1
class Program{
  public void main(String args []){
    System.out.println("Hello World");
  }
}
\end{verbatim}

Unlike Java, C\# does not require a class construct for error-free
parsing and compilation, and thus this repair was only applied to Java
snippets.

\subsubsection{Repair 2 - Semicolon}

Java and C\# statements require a semicolon (``;") at the end in
order to parse and compile correctly. To decide whether a ``;" should
be added to a statement, we run a set of heuristic checks on each line
of the snippet; we add the semicolon if all the following conditions are true:
\begin{enumerate}
  \item If the line does not contain any of the tokens \texttt{;}, \texttt{\{}, \texttt{(}, and \item if the line does not contain any of the tokens \texttt{class}, \texttt{if}, \texttt{else}, \texttt{do}, \texttt{while}, \texttt{for}, \texttt{try}, \texttt{catch}, and  
  \item if the line does not end with the tokens \texttt{$=$} and \texttt{\}}. 
 \end{enumerate} 

With check 1, we avoid double-adding ``;" and avoid adding a ``;"
at the line of an opening brace, before the opening brace has been
closed. With check 2 and 3 we avoid corrupting originally parsable code. With check 2 we avoid the following situation:
\begin{verbatim}
\\Repair 2
try; <--  will corrupt
{  <code>
}catch...
\end{verbatim}
With check 3 we avoid the following situations: 
\begin{verbatim}
\\Repair 2 inside assignment
 Double s_dev = ; <-- will corrupt
     Math.pow(sum(mean_sq(al))/al.size(),0.5);
\\Repair 2 between if-else
if ()  
{   <code>
} ; <-- will corrupt
else
{   <code>
}
\end{verbatim}

\subsubsection{Study Workflow for Repairs}

\begin{figure}
	\centering
	\includegraphics[width=0.5\textwidth]{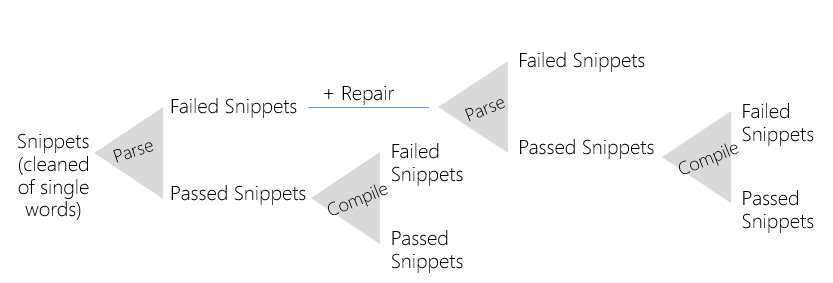} 
	\caption{Sequence of operations while applying repairs}
	\label{fig:opseqfx}
\end{figure}

The workflow for studying the effect of repairs is similar to the one
used for the initial parsing/compiling processes (as shown in
Figure~\ref{fig:opseq}), except that now we have also incorporated
repairs. The process is depicted in Figure~\ref{fig:opseqfx}.

Like the workflow of Figure~\ref{fig:opseq}, here at each stage we
only compile snippets that have passed the prior step. Failed snippets are
repaired and parsed again. The snippets that succeed at parsing are in turn
compiled. For Java, we carry out two repairs sequentially, whereas for
C\# one repair is applied.

\subsubsection{Results after Repairs}
\label{subsubsec:respostfx}

Table~\ref{tab:res_postFix} shows the parsing and compilation results
obtained for C\# and Java after repairing the snippets. Again, the
numbers here reflect the collection of non-single-word snippets.

\begin{table}
\centering
\caption{Summary of results for C\# and Java snippets after repairs}
\label{tab:res_postFix}
\begin{tabular}{lll}
                                                               & \textbf{C\#}      & \textbf{Java}     \\ \hline
\begin{tabular}[c]{@{}l@{}}\textbf{Total snippets}\\      \textbf{after removal}\end{tabular}                & 514,992           & 572,742           \\ \hline
\begin{tabular}[c]{@{}l@{}}\textbf{Parsable snippets}\\      \textbf{after repairs}\end{tabular}   & 135,421 (26.30\%) & 110,203 (19.24\%) \\ \hline
\begin{tabular}[c]{@{}l@{}}\textbf{Compilable snippets}\\      \textbf{after repairs}\end{tabular} & 986 (0.19\%)        & 17,286 (3.02\%)  
\end{tabular}
\end{table}

Although the repairs did not significantly increase the usability
rates for C\#, the improvements were quite significant for parsing
Java snippets. The parse rate of C\# improved by only 1.12\% (from 25.18\% to 26.30\%), whereas
for Java the improvement was 13.02\% (from 6.22\% to 19.24\%). The compilation rate did not
change for C\#, whereas for Java it improved by 1.42\% (from 1.6\% to 3.02\%). There's a significant improvement 
on the parsable rate of Java.

Again, our approaches are heuristic, and may break some previously parsable or compilable snippets.
But we can still see an increase in usability rates.

Even though the parsing and compilation rates improved for Java, the
number of usable snippets is still one order of magnitude lower than
the numbers for JavaScript and Python.

\section{Qualitative Analysis}
\label{sec:qualitative}

\begin{figure}
	\centering
	\includegraphics[width=0.5\textwidth]{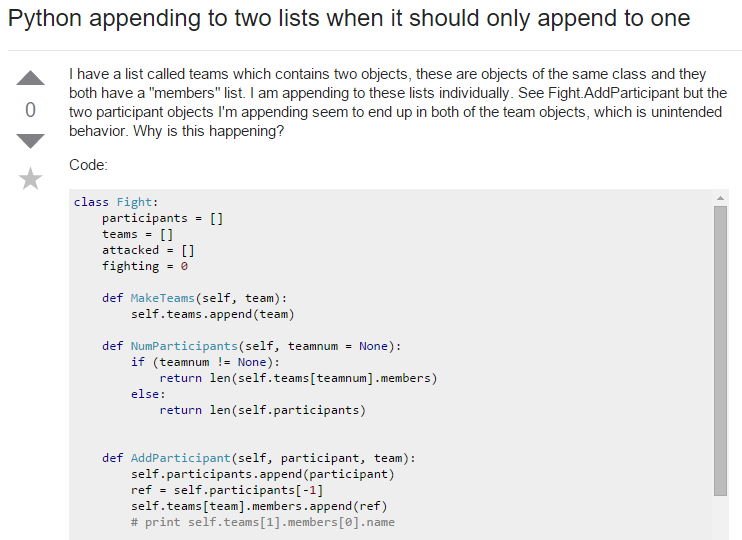} 	
	
	\vspace{0.2 in}
	
	\includegraphics[width=0.5\textwidth]{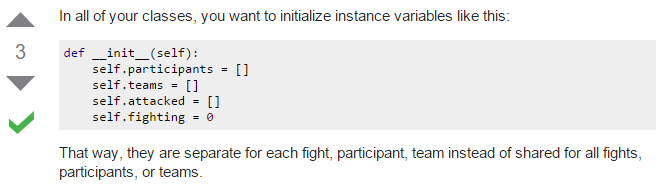}
	\caption{Example of an incomplete answer in Stack Overflow}
	\label{fig:incompleteAnswer}
\end{figure}

	
	

Based on the usability and popularity, we choose Python as the target
language for further analysis. In order to investigate whether the
runnable Python snippets can answer the questions {\em correctly} and
{\em completely}, we perform a 3-step qualitative analysis on randomly
selected snippets. By {\em
  correctness}, we mean the snippet giving a concise solution to the
question; for specific coding questions with bugs, as in
Figure~\ref{fig:incompleteAnswer}, the answer is {\em correct} if it
points out the erroneous part and fixes particular lines of code.
By {\em completeness}, we mean that the snippet
itself is a full answer to the question; we do not need to add any
additional code to answer the question.
Figure~\ref{fig:incompleteAnswer} is an example of correct but
incomplete answer, the snippet fixes the bug in the original code but
we have to mix the question and answer snippets to get the full
answer. 

The 3-step qualitative analysis is as following:

\noindent
{\bf Step 1}

\noindent

\begin{table}
\caption{Features used to assess the quality of the snippets.}
\label{tab:features}
\begin{tabular}{rl}\\
  1. & Votes for the question \\
  2. & Votes for accepted answer\\
  3. & Total number of answers \\
  4. & Is the accepted answer also the best answer?\\
  5. & Questioner's reputation score\\
  6. & Answerer's reputation score\\
  7. & Does the title correctly summarize the question \\
     & described? \\
  8. & Is the question's description clear?\\
  9. & Is it a specific coding question?\\
  10. & Does the snippet answer the question correctly \\
      & and completely?\\
  11. & Is it a single word snippet?\\
  12. & Is it a single line snippet?\\
  13. & Is there any surrounding context/explanation?\\
  14. & Number of comments\\
  15. & Is there any questioner's compliment in \\
      & comments?\\
  16. & Question's tags\\
\end{tabular}
\end{table}

We randomly chose 50 runnable Python snippets. For each snippet, we
investigate the features listed in Table~\ref{tab:features}.
We found out that the proportion of snippets that answer the question
is low (16\%). We discovered a strong correlation between single word
snippets and snippets answering the question, that is, among the 50
selected snippets, none of single word snippets answer the question,
and all of the snippets that answer the question are non-single word
snippets. The proportion of single word snippets is 64\%. \\

\noindent
{\bf Step 2}

Based on the results of Step 1, we removed single word
snippets from all runnable Python snippets, and then randomly chose
another 50 snippets. We investigated the same aspects as in Step 1,
except for No.11 (Is it a single word snippet?).

After removing single word snippets, the proportion of snippets that
answer the question increases to 44\%.  From Step 2, we discovered
another negative correlation between single line snippets and snippets
answering the question. Among the 21 snippets that answer the
question, 19 are multiple line snippets, and among the 20 single line
snippets, only 2 answer the question.\\

\noindent
{\bf Step 3}

Finally, we removed the single line snippets, and chose 50 snippets
randomly again. We investigated the same aspects as in Step 1, except
for No.11 (Is it a single word snippet?) and No.12 (Is it a single
line snippet).
Again, the proportion of snippets that answer the question increases,
to 66\%. Moreover, for the 17 snippets that do not answer the
question, 12 of them are incomplete, but correct, answers.

From the result of 3-step qualitative analysis, we can see that
multiple-line snippets can best answer the questions. This subset
contains 40,245 runnable snippets (29.8\% of all runnable Python
snippets).

\section{Google Search Results}
\label{sec:googleResults}

In this section, we explore the overlap between Google search results
and the usable Python snippets. Specifically, we check if the top
results from Google for several queries contain parsable or runnable
snippets, as well as these snippets' overall quality.

The methodology was as follows. We selected 100 programming related
questions from SO's highest voted questions about Python, and use them as queries
using the Google search API. We add the constraint
``site:stackoverflow.com'' and the keyword ``Python'' in the in the
queries. Moreover, because our database was downloaded in April 2014,
we also add a date range restriction.

\begin{table}
\caption{Usability Rates of Top Results from Google}
\label{tab:usabilityGoogle}
\centering
\begin{tabular}{ccc}
\hline
\textbf{} & \multicolumn{1}{l}{Parsable} & \multicolumn{1}{l}{Runnable}\\ \hline
Top 1              & 78.1\%                        & 30.8\%                             \\ \hline
Top 10            & 77.9\%                        & 29.3\%                             \\ \hline
\end{tabular}
\end{table}

The accepted answers' usability rates of the Top 1 and Top 10 results from Google are
shown in Table~\ref{tab:usabilityGoogle}. They are 
high. As described in Section~\ref{subsec:findings}, we had found that
the usability rates of all the Python snippets in SO are 76\% parsable
and 25\% runnable. The top results on 100 queries to Google on the
same SO data have usability rates above those averages. Moreover, the
Top 1 results have an even higher usability rate than the Top 10
results.

Also, we find that 33.7\% of Top 1 results and 32.5\% of Top 10 results
are multiple line snippets. Both higher than the average of 30\%.  So,
both from our usability perspective and qualitative analysis
perspective, the Google Top 10 search results are better than average, and
the Top 1 results are the best.

\begin{figure}
\centering
\includegraphics[width = 0.5\textwidth]{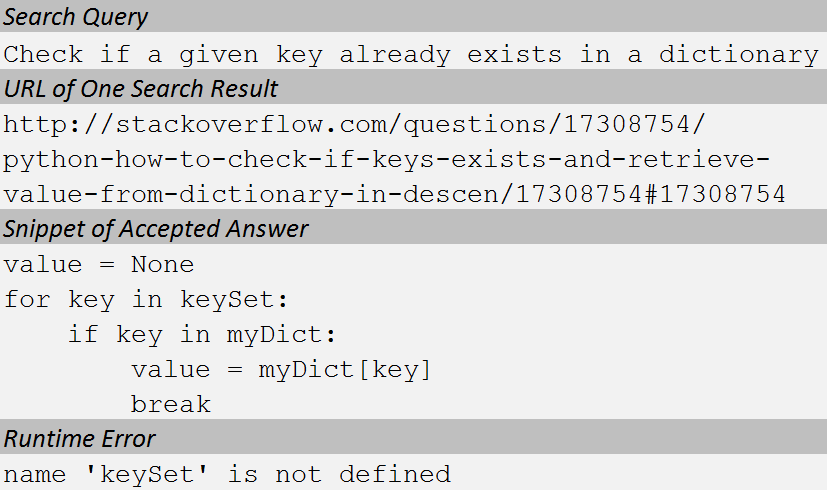}
\caption{Example of Google Search Result}
\label{fig:queryExample}
\end{figure}

From the results above, we can also see that the Google top results
have a low runnable rate, although higher than average. The main
problems encountered in the parsable but not runnable snippets from
Google results are those already described for the entire SO snippets
(see Section~\ref{subsec:errMsg}). Specifically, the majority of them
suffered from undefined names or modules. An example of a parsable but
not runnable Google search result to the query ``Check if a given key already exists in a dictionary'' is shown in Figure~\ref{fig:queryExample}.

Expecting snippets to be runnable as-is may be too strong of a
constraint. Parsable snippets seem to be a much more fertile groud as
the base for future automatic code generation. Given our analysis of
the causes of runtime errors, it seems it should be possible to repair
a large percentage of them automatically. For example for Python,
missing symbol names often indicate a piece of information that needs
to come from elsewhere -- another snippet, or some default initialization.

Note that we used the questions as-is as queries for Google; not
surprisingly, Google always returned those SO questions as the Top 10
hits in each query.  Out of the 100 queries we selected, 85 original
ones were returned as the first hit by Google. Although 15 original
links were not ranked as Top 1, 12 of them were in Top 10. The reason
for them not being the first one is that Google seems to have a
special heuristic to dealing with ``daterange'' restrictions.  If we
remove the ``daterange'' restriction in our search query, the original
ones will appear in Top 1. However, 3 out of 100 queries were not in
Top 10 list by Google. We looked at these three cases: one is because
of the date range restriction, the second one is because it is a new
query out of our date range, and the last one seems to genuinely be
because of Google ranking algorithms.  

The very high hit rate and, in particular, the top 1 results, confirm
Google's efficiency in retrieving relevant information from the Web,
something that our work leverages, by design. However, the usability
rates on the top hits were encouragingly high, and that is orthogonal
to Google's efficiency in finding the most relevant results. These
higher-than-average usability rates may be because we used the most
popular queries; users of SO value complete answers that have good
code snippets, so it is not surprising that the most popular queries
have snippets that are better than average.

In general, users search using words that are not exactly the
same as the words in the SO questions, so the best snippet of code for
their needs may not be in the first position; but, as it is usually
the case with Google, it is likely in the top 10 positions. The
usability of the snippets in the top 10 positions were not as high, but they
were still very high (78\% parsable, 29\% runnable), and
above average of the entire set of Python snippets.

The Google search results over SO snippets are very encouraging. They
show that it is possible to go from informal queries in natural
language to relatively usable, and correct, code in a large percentage
of cases, opening the door to the old idea of programming environments
that ``do what I mean''~\cite{pilot:1966}. This is possible now due to the
emergence of very large mixed-language knowledge bases such as SO.

\section{Related Work}
\label{sec:back}
Various studies have been done on SO, but focus primarily on user
behavior and their interactions with one another. These works made
attempts at identifying correlations between different traits of SO
users. For example ~\cite{Morrison} showed a correlation between the
age and reputation of a user by exploring hypotheses such as the fact
that older users having a bigger knowledge of more technologies and
services. Shaowei et al.~\cite{Shaowei} provided an empirical study on
the interactions of developers in SO, revealing statistics on
developers' questioning and answering habits. For instance, they found
that a few developers ask and answer many questions. This social
research might be important for our prioritization of snippets of
code.
 
Among the works that utilize code available in the public domain for
enhancing development is that of Wong et al.~\cite{Wong}. They devised
a tool that automatically generates comments for software projects by
searching for accompanying comments to SO code that are similar to the
project code. They did so by relying on clone detection, but never
tried to actually use the snippets of code. This work is very similar
to Ponzanelli et al.~\cite{Ponzanelli} in terms of the approach
adopted. Both mine for SO code snippets that are clones to a snippet
in the client system, but Ponzanelli et al.'s goal was to integrate SO
into an Integrated Development Environment (IDE) and seamlessly obtain
code prompts from SO when coding.

Ponzanelli was involved in another work~\cite{Bacchelli}, where
they presented an Eclipse plugin, Seahawk, that also integrates SO
within the IDE. It can add support to code by linking files to SO
discussions, and can also generate comments to IDE code. Similarly,
Thummalapenta et al.~\cite{thummalapenta} present a tool called
PARSEWeb that assists in reusing open source frameworks or libraries
by providing an efficient means for retrieving them from open source
code. 

With regards to assessing the usability of code, our central
motivation, our study comes close to Nasehi et al.'s
work~\cite{Seyed}. They also analyzed SO code with the motivation of
finding out how easy it is to reuse it. In particular, they delved
into finding the characteristics of a good example. The difference for
our approach was their criteria for assessing the usability of the
code. They adopted a holistic approach and analyzed the
characteristics of high voted answers and low voted answers. They
enlisted traits related to a wide range of attributes of the answers
by analyzing both the code and the contextual information. They looked
into the overall organization of the answer - the number of code
blocks used in the answer, the conciseness of the code, the presence
of links to other resources, the presence of alternate solutions, code
comments, etc. The execution behavior of the code was not among their
usability criteria.

Semi-automatic or automatic programming, a development realm towards
which this work takes an initial step, has also been explored in
different ways by software practitioners. For instance, Budinskey et
al.~\cite{Budinskey} and Frederick et al.~\cite{Frederick}, analyze
how design patterns could assist in automatically generating software
code. Other similar works include~\cite{Noble:2002:PS:646159.680030},
\cite{Zheng:2011:AM:1985793.1986010}, and
\cite{Ohtsuki:1999:SCG:518898.785728}. The similarity in these works
is that structured abstractions of code provide a good indicator about
the actual implementation. They built tools that exploit this
narrative and generate implementations from design patterns. Such a
strategy would be highly challenging to use when trying to reproduce
usable SO code (with respect to compilation), as those codes are
usually small snippets that do not follow familiar patterns.

\section{Discussion and Conclusion}
\label{sec:concl}

Some of our experiment choices deserve an explanation:

\begin{itemize}
\item In SO, the concepts of accepted answer and best answer are
  different. Accepted answer is the one approved by the questioner,
  while best answer is voted by all viewers. We chose accepted answer
  in this work because we believe that in a Question\&Answer forum as
  SO, the questioner the one who has the best judgement of whether the
  answer solves the problem. However, it is possible that the
  questioner makes mistakes and that the answer voted the best by
  other viewers is most usable. In the future we will evaluate the
  usability of best answers and compare the results with those of
  presented here.

\item Our definition of usability is purely technical, and does not
  include the concept of usefulness other than indirectly, by the fact
  that the analyzed snippets are in the accepted answers. It is
  possible that a snippet that does not parse is more useful than the
  one that runs; or that a snippet that does not parse or run is still
  useful to the answer the question. For example, if the question
  asked in SO is not a specific programming query, people may answer
  with pseudo code, which is not usable in our case, but may also
  answer the original question. Those cases, however, will always be
  out of reach of automatic tools, as they will require many more
  repairs or even translation from pseudo-code to actual code. As
  such, this study focused conservatively on those snippets that are
  part of accepted answers and that show good potential to being used
  as-is or with little repairs.

\end{itemize}

In this paper, we examined the usability of code snippets in
Stack Overflow. The purpose of our usability analysis is to understand
the extent to which human-written snippets of code in sites like SO
could be used as basic blocks for automatic program generation.
We analyzed code snippets from all the accepted answers for four
popular programming languages. For
the two statically-typed, compiled languages, C\# and Java, we
performed parsing and compilation, and for the two dynamic languages,
Python and JavaScript, we performed parsing and running
experiments. 
The results show that usability rates for the two dynamic languages is
substantially higher than that of the two statically-typed, compiled
languages. Heuristic repairs improved the results for
Java, but not for C\#. Even after the repairs, the compilable rates
for both Java and C\# are very low. The results lead us to believe
that Python and JavaScript are the best choices for program synthesis
explorations.

Usability as-is, however, is not enough to ensure that the snippets
have high information quality. Our qualitative analysis on the most
usable snippets showed that multiple line snippets have the highest
potential to answer the questions. We found 40K+ of these for Python,
meaning that there is a good potential for processing them
automatically.

Finally, in order to close the circle on our original vision, we
investigated the extent to which the top results of queries on SO
using Google Web search contain these usable snippets. The results are
very encouraging, and show a viable path from informal queries to
usable code.


\section{Acknowledgments}
This work was partially supported by grant No. 1218228 from the
National Science Foundation and by a grant from the DARPA MUSE
program. 

%
\bibliographystyle{abbrv}
\bibliography{MSR2016}  
%
%

\end{document}